\documentclass[aps,prl,twocolumn]{revtex4-1}

\usepackage[dvips]{graphicx}
\usepackage{amssymb,amsfonts,amsmath}
\usepackage{color}
\usepackage{colortbl}
\usepackage{wasysym}
\usepackage{ifsym}

\begin{document}
%\preprint{}
\title{Graphical notation reveals topological stability criteria for collective dynamics in complex networks}
\author{Anne--Ly Do,$^1$ Stefano Boccaletti,$^2$ and Thilo Gross $^3$}
\email[]{ly@mpipks-dresden.mpg.de}
\affiliation{\mbox{$^1$ Max-Planck-Institute for the Physics of Complex Systems, Dresden, Germany}\\
\mbox{$^2$ Technical University of Madrid, Center for Biomedical Technologies, Madrid, Spain}\\
\mbox{$^3$ University of Bristol, Merchant Venturers School of Engineering, Bristol, UK}}

\date{\today}

\begin{abstract}
We propose a graphical notation by which certain spectral properties of complex systems can be rewritten concisely and interpreted topologically.
Applying this notation to analyze the stability of a class of networks of coupled dynamical units, we reveal stability criteria on all scales.
In particular, we show that in systems such as the Kuramoto model the Coates graph of the Jacobian matrix must contain a spanning tree of positive elements for the system to be locally stable.  
\end{abstract}

\maketitle
Discovering how the interactions between constituents of a system determine its macroscopic behavior is a central aim of physics. 
Ideally, we find properties in the system's detailed organization that have direct system-level implications.
Significant progress has been made in identifying such properties on a local level. 
For instance the implications of the degree distribution, the distribution of a network's links among the nodes, is well understood \cite{Newman}. 
By contrast, identifying meso-scale properties that have a distinct effect on macroscopic behavior, despite recent advances \cite{Fortunato}, remains a difficult challenge.
A starting point is often to link the macro-behavior to spectral properties of certain matrices, such as the adjacency, the different graph Laplacians, or the Jacobian matrix of a dynamical system.     
Here again, the dependence on local properties, i.e., certain rows or diagonal elements, can be discovered with relative ease. 
By contrast, analytical insights on how meso-scale patterns affect the spectrum are difficult to obtain, as they typically require the 
evaluation of determinants leading to complex expressions. 

In this Letter we propose a graphical notation, reminiscent of Feynman diagrams, that facilitates computing spectral implications of meso- and macro-scale structures.
We illustrate the usage of the notation by investigating dynamical stability of stationary and phase-locked solutions in a class of symmetrically coupled dynamical systems, containing for instance the Kuramoto model \cite{Kuramoto,Acebron}. 
Here, the proposed notation greatly reduces the complexity of the mathematical expressions and enables the derivation of topological stability criteria on all scales. 
In particular, we show that in the systems under consideration the graph defined by the off-diagonal elements of the Jacobian must contain a spanning tree of positive links to admit stable solutions.

Any given matrix ${\rm \bf J}$ can be interpreted as defining the connectivity of an abstract weighted network $\mathcal{G}$, the so-called Coates graph. 
In this network two nodes $i,j$ are connected if $J_{ij} \neq 0$. 
We can interpret a product of different matrix elements as defining a (not necessarily connected) network motif, e.g.{} $J_{ij}J_{jk}J_{ki}$ corresponds to a triangle path. 
The advantage of this graphical reading is that complex structures appearing in systems of equations can be described using 
network terminology. 

\begin{figure}
\includegraphics[width=0.48\textwidth]{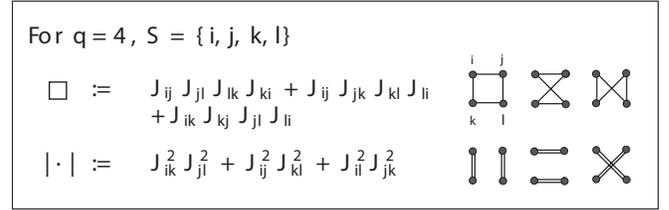} 
\caption{Examples for the graphical notation. Symbols denote the sum over all non-equivalent possibilities to build the depicted subgraph with the $q$ nodes $\in S$. Plotted are two example terms and their algebraic and topological equivalents.}
\label{Examples}
\end{figure}

It is useful to define a basis of symbols, $\times, |, \triangle, \Square, \pentagon, \ldots$, denoting sums over all non-equivalent cycles of length $n=1,2,3,4,5,\ldots$, respectively.
We allow concatenation of these symbols by the `$\cdot$' sign.
For a given set $S$ of node indices, products of symbols denote the sum over all non-equivalent possibilities to realize the depicted motif with the   
nodes in $S$. For instance, if $S=\left\{i,j,k,l,m\right\}$ then $\times \cdot \square$ represents the sum over all 
products of elements that would be read as a cycle containing four nodes and a self-loop,
e.g. $J_{ii}J_{jk}J_{kl}J_{lm}J_{mj}$ (cf.~Fig.~\ref{Examples}).
Finally, we define $\Phi_{S}$ as the sum over all products corresponding to all acyclic graphs that can be drawn by placing $\left|S\right|$ (undirected) links such that each link starts at a different node in $S$.  
 
In the remainder of this Letter we illustrate the usage of the proposed notation by considering the synchronization of dynamical units. 
This subject was chosen as it is presently of broad interest in physics and appears in many fields including biology, ecology, and engineering \cite{Kurths,Boccaletti,Arenas}. The paradigmatic model proposed by Kuramoto \cite{Kuramoto} opened the field for detailed studies of the interplay between the structure of the interaction network and collective phenomena \cite{Mirollo,Wu,Pecora,Kawamura}. These revealed the influence of various topological measures, such as the clustering coefficient, the diameter, and the degree or weight distribution, on the propensity to synchronize \cite{Chavez,Nishikawa2006,Lodato}. 
However, recent results \cite{Arenas,Nishikawa,Mori} indicate that beside global topological measures also details of the exact local configuration can crucially affect synchronization. 
This highlights synchronization of phase oscillators as a promising example in which it may be possible to understand the interplay between local, global, and mesoscale constraints on stability, that severely limit the operation of complex technical and institutional systems \cite{Motter,Havlin}.

To provide a specific example we focus on the Kuramoto model \cite{Acebron}, describing a network of $N$ bidirectionally coupled phase oscillators 
\begin{equation}\label{ODE}
\dot{x_i}=\omega_i+\sum_{j\neq i} A_{ij}\sin(x_j-x_i)\ , \quad \forall i\in 1\ldots N	\ ,
\end{equation}
where $x_i$ and $\omega_i$ are the phase and intrinsic frequency of oscillator $i$ and ${\rm \bf A}$ is the weight matrix of an undirected,  network. 
The model can exhibit phase-locked states, which in a co-rotating frame correspond to steady states of \eqref{ODE}. 
The local stability of such states is determined by the eigenvalues of the Jacobian matrix ${\rm \bf J} \in\mathbb{R}^{N\times N}$ defined by $J_{ik}=\partial\dot{x_{i}}/\partial x_{k}$. If all eigenvalues of ${\rm \bf J}$ are negative, then the state under consideration is locally stable.

For deriving topological stability criteria we apply Jacobi's signature criterion (JSC), also called Sylvester criterion. 
The JSC states that a hermitian matrix ${\rm \bf J}$ with rank $r$ is negative definite iff all principal minors of order $q\leq r$ have the sign of $(-1)^q$ \cite{LiaoYu}, i.e., iff 
\begin{equation}\label{stability_cond}
	{\rm sgn}\left(D_{\left|S\right|,S}\right)=(-1)^{\left|S\right|}  \ \forall \  S\subset\left\{1,\ldots,N\right\}, \ \left|S\right|\leq r 
\end{equation}
where $D_{\left|S\right|,S}:=\det\left(J_{ik}\right)$, $i,k\in S$. Note that the whole set of equations~\eqref{stability_cond} constitutes a sufficient condition for stability, whereas each equation already constitutes a necessary condition. 
 
Stability analysis by means of JSC is used in control theory \cite{LiaoYu} and has been applied to problems of different fields from fluid- and thermodynamics to offshore engineering and social networks \cite{Beckers,Soldatova,CaiWuChen,DoRudolfGross}. 
An interesting property of the JSC is that it provides necessary conditions on all scales. 
However, application of the criterion were previously limited mostly to systems with few degrees of freedom because of the difficulties associated with computing large determinants. 

In the following we rewrite the JSC conditions in the notation proposed above. 
By direct comparison we find 
\begin{eqnarray*}
D_{1,S}&=&\times\\
D_{2,S}&=&\times \cdot \times -| \\
D_{3,S}&=&\times \cdot \times \cdot \times - \times \cdot  | +2\triangle\\
D_{4,S}&=&\times \cdot \times\cdot \times\cdot \times -\times\cdot\times \cdot |+|\cdot|+2\times \cdot \triangle
- 2\Square
\end{eqnarray*}
and generally 
\begin{equation}\label{formationrule}
	D_{\left|S\right|,S}=\sum{\tiny \text{all combinations of symbols with $\sum n=\left|S\right|$}},
\end{equation}
where symbols with $n>2$ appear with a factor of $2$ that reflects the two possible orientations in which the corresponding subgraphs can be paced out.
Symbols with an even (odd) number of links carry a negative (positive) sign related to the sign of the respective index permutation in the Leibniz formula for determinants \cite{LA}. We note that for instance that $D_{6,S}$ has only 11 terms in the proposed graphical notation in contrast to 720 terms in conventional notation.

\begin{figure}
\includegraphics[width=0.47\textwidth]{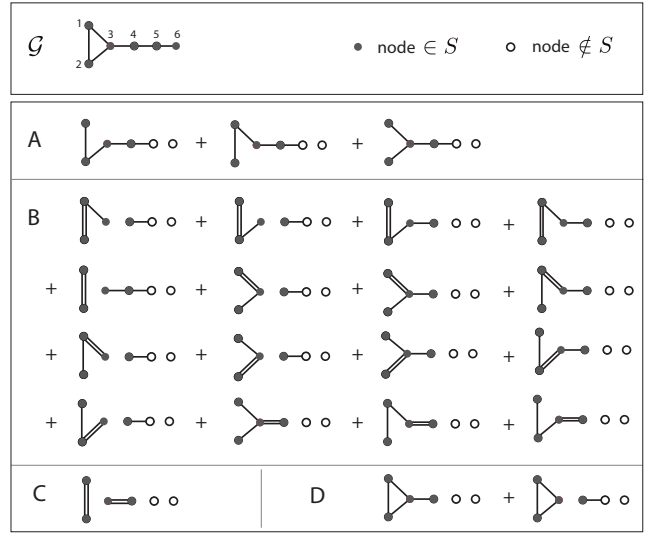}
\caption{Symbolic calculation of a determinant using the graphical notation. Consider the minor $D_{4,S}$ of the graph $\mathcal{G}$ sketched above. If $S$ is chosen to be the set of nodes plotted in grey, the terms of the fourth JSC, $D_{4,S}$, can be written as $\times\cdot\times\cdot\times\cdot\times=J_{11}J_{22}J_{33}J_{44}=(-1)^4(J_{12}+J_{13})(J_{21}+J_{23})(J_{31}+J_{32}+J_{34})(J_{43}+J_{45})=A+B+C+2D$, $-\times\cdot\times \cdot |=-(B+2C)$, $|\cdot|= C$, $2\times \cdot \triangle= -2D$ and $-2\Square= 0$. We can immediately read off that $D_{4,S}\equiv A = \Phi_{S}$, defined in the text. \label{Subgraphs}}
\end{figure}

The highly aggregated but somewhat awkward Eq.~\eqref{formationrule} applies to all systems of symmetrically coupled one-dimensional units.
A more convenient expression can be obtained by focusing specifically on systems having a zero row-sum (ZRS), such that $J_{ii}=-\sum_{j\neq i}J_{ij}$.
In the Kuramoto model (and many other systems), the ZRS results from a force balance along the links.  
In the topological reading, it allows replacing all self-loops $J_{ii}$ by the negative sum over all links connected to respective node $i$.
For the first term, $\times^{\left|S\right|}$, of a minor $D_{\left|S\right|, S}$ this leads to a sum over all graphs that 
can be drawn by placing $\left|S\right|$ (undirected) links such that every link starts from a distinct node in $S$.  
By elementary combinatorics it can be verified that all other terms of $D_{\left|S\right|, S}$ cancel exactly those subgraphs in $\times^{\left|S\right|}$ that contain cycles (cf. Fig.~2), leaving exactly $\Phi_{S}$ defined above. 
For systems with ZRS, we can thus express the minors as 
\begin{equation}\label{det_n_2}
	D_{\left|S\right|, S}=(-1)^{\left|S\right|}\Phi_{S}.
\end{equation}
Thus the JSC stability conditions translate to 
\begin{equation}\label{stability_cond2}
	\Phi_{S}>0, \quad \forall \ S \ \text{with} \ \left|S\right|\leq r \ .
\end{equation}
We remark that $\Phi_{S}$ with $\left|S\right|=N-1$ is the sum over all spanning trees of the network, such that Kirchhoff's Theorem \cite{Schnakenberg, Li} appears as the special case of Eq.~\eqref{det_n_2}. 

The graphical stability conditions $\Phi_{S}>0$ conveniently conceals the complexity of the underlying determinants. 
Below we show that results can now be obtained by reasoning on the graphical level, i.e., without digging up the complex underlying expressions. %For instance, $J_{ij}>0$ for all $i,j$ is a sufficient condition for stability because $\Phi_S>0$ for all $S$.  
 
Consider a network containing a tree-like branch that is only connected to the rest of the network by a single 
link. We choose $S$ as the set of nodes that are located in the branch and focus on the condition $\Phi_{S}>0$.
The rules for constructing the network motifs in $\Phi_{S}$ imply they must contain one link starting from every node in $S$. By starting from the nodes of degree 1 (having only one link) and working downward one can see that there is only one motif contributing to $\Phi_{S}$, which consists of all links in the branch and the link connecting it to the rest of the network. The condition $\Phi_{S}>0$ therefore implies that the number of links in $\Phi_S$ associated with a negative elements must be even. Further, if this number were greater than zero, one could always find a part of the branch to which the same conditions apply, but which contains only one of the negative links, such that a stability condition on a smaller scale is violated. Thus, stability requires that all links appearing in such tree-like branches correspond to positive entries of ${\rm \bf J}$.

Using similar arguments as above the implications of different meso-scale motifs can be determined.
The analysis reveals restrictions on (a) the number and position of potential negative links and (b) the absolute value of their weights. We find that restrictions of type (a) can be subsumed under one general condition: To admit stable solutions, a dynamical system with symmetric Jacobian and ZRS must possess a spanning tree made up entirely of positive elements.

To prove the statement above consider that in a network without a positive spanning tree it must be possible to partition the nodes into two nonempty sets $I_1$, $I_2$ such that  
\begin{equation} \label{negativecut}
	J_{ij}\leq 0 \quad \forall \ i,j \ | \  i\in I_1,\  j\in I_2  \ .
\end{equation}
The idea is now to evaluate the stability conditions~\eqref{stability_cond2} for different $S\subseteq I_1$ thereby exploiting that all links leading out of $I_1$ have negative weights. 
It is convenient to define $E^*$ as the set of links connecting $I_1$ and $I_2$, and $X=\left\{x_{1},\ldots,x_m\right\}$ as the subset of nodes $\in I_1$ incident to at least one link from $E^*$ (`boundary nodes'). 
Further, we define $\sigma_i$ as the sum over all elements of $E^*$ incident to $x_i$, and, for any subset $Y$ of $X$, $\sigma_{Y}:=\prod_{m\in Y}\sigma_m$. 
Finally, for any subset $Y$ of $X$, we define $\tau_{Y}$ as the sum over all forests of $\mathcal{G}$ that (i) span $I_1$, and (ii) consists of $\left|Y\right|$ trees each of which contains exactly one element from $Y$. 
We can now write
\begin{equation}\label{expansion}
	\Phi_{I_1\setminus C}=\sum_{B\subseteq X\setminus C}\sigma_B\tau_{B\cup C} \ ,
\end{equation} 
where $B$ and $C$ are disjoint subsets of $X$. We show that this is incompatible with the stability condition \eqref{stability_cond2} by the contradiction
\begin{align}
	\sum_{C\subseteq X}\underbrace{(-1)^{\left|C\right|}\sigma_{C}}_{\substack{>0\ \text{per} \\\text{construction}}}\underbrace{\Phi_{I_1\setminus C}}_{>0}&=\sum_{C\subseteq X}(-1)^{\left|C\right|}\sigma_{C}\sum_{B\subseteq X\setminus C}\sigma_B\tau_{B\cup C} \nonumber\\
	&=\sum_{\substack{C\subseteq X\\B\subseteq X\setminus C}} (-1)^{\left|C\right|}\sigma_{B\cup C} \tau_{B\cup C}\nonumber\\
	&=\sum_{\substack{A\subseteq X\\C\subseteq A}} (-1)^{\left|C\right|}\sigma_{A} \tau_{A}\nonumber\\
	&=\sum_{A\subseteq X}\sigma_{A} \tau_{A}\underbrace{\sum_{C\subseteq A}(-1)^{\left|C\right|}}_{=0}=0 \ . \nonumber
\end{align}
Therefore the existence of a spanning tree of positive elements is a necessary condition for stability. 

Note that the (global) spanning tree criterion has distinct implications for meso-scale properties of $\mathcal{G}$. Any unbranched path can maximally contain one negative link, and the number of negative links in the network is limited by the number of independent cycles (Fig~3). 

Let us now turn to stability conditions of type (b), which restrict the absolute value of potential negative links and result from smaller scale stability conditions:
Consider for instance an unbranched segment of a cycle of $\mathcal{G}$ that consists of $d$ links $c_i$ with weights $w_i$, one of which, say $w_x$, is negative.
Evaluating Eq.~\eqref{stability_cond2} for a series of sets $S$ $\subset \left\{1,\ldots,d\right\}$, we find that stability requires
	\[	
	\left|w_{x}\right|<\frac{\prod_{i\in I^*}w_i }{\sum \text{ \small all distinct products of $(d-2)$ factors $w_i$, $i\in I^*$}} ,
\]
where $I^*= \left\{1,\ldots,d\right\}\setminus x$.

\begin{figure}
\begin{center}
\includegraphics[width=0.37\textwidth]{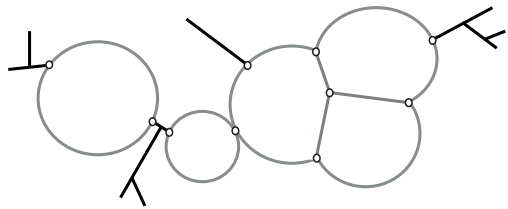}
\end{center}
\caption{Decomposition of a graph $\mathcal{G}$ in acyclic parts (black), unbranched segments of cycles (grey) and branching points (open black). 
Lines represent paths of $\mathcal{G}$ that may contain several nodes.
Stability requires that (i) the acyclic parts only contain links with positive weights; (ii) any unbranched segment of a cycle contains at most one link with negative weight; (iii) the number of links with negative weight does not exceed the number of independent cycles (here, 5). 
\label{Decomposition}}
\end{figure}

% KURAMOTO IMPLICATIONS
Applied to the Kuramoto model, the criterion derived above implies that in any phase locked state a spanning tree must exist on which the 
phase difference between any two coupled oscillators obeys $|x_j-x_i|<\pi/2$. In networks where this condition is violated the criterion 
points to local interventions that enhance synchronizability.  

%%% OTHER APPS %%%
We note that the results derived above are not contingent on the specific form of the Kuramoto model. They apply to all symmetrically coupled systems obeying the ZRS condition, which includes all systems of the form
\begin{equation} 
\label{generalization} 
\dot{x_i}=C_i+\sum_{j\neq i} A_{ij}\cdot O_{ij}(x_j-x_i)\ , \quad \forall i\in 1\ldots N	\ , 
\end{equation}
where the $A_{ij}$ are the weights of a symmetric coupling matrix and $O_{ij}$ are odd functions.
Besides systems of coupled oscillators, Eq.~\eqref{generalization} can for instance describe a range of variants of the Deffuant model of social opinion formation and ecological meta-population models. 
We emphasize that the derived results remain valid for heterogeneous networks containing different $A_{ij}$, $O_{ij}$, and $C_i$. 

%%%%%%%%%%%%%%%%%%%%%%%%%%%
Although the ZRS simplifies the derivations above, similar calculations can be carried out for systems violating the ZRS condition, e.g.{} \cite{DoRudolfGross}. We applied the graphical notation to an adaptive extension of the Kuramoto model, where the ZRS is violated and the connection strength coevolves with the dynamics on the network according to ${\rm d/dt}(A_{ij})=\cos(x_j-x_i)-b\cdot A_{ij}$. Such systems have recently received much interest in physics because of their role in neuroscience \cite{ItoKaneko,ZhouKurths,Almendral}. Because of space constraints we postpone the detailed discussion of the model to a subsequent publication, but note that proceeding similarly as above leads to a stronger spanning tree condition for the adaptive system: Stability requires in this case the existence of a spanning tree where linked nodes obey $|x_j-x_i|<\pi/4$.  
 
The proposed notation is also applicable to certain questions not pertaining to dynamics, for instance to the question of isospectrality of hermitian matrices \cite{Eskinetal,Shirokov}.
The key idea is that the characteristic polynomial $\chi$ of a hermitian matrix ${\rm\bf A}\in\mathbb{C}^{n\times n}$ can be expressed as $\chi(\lambda)=D_n({\rm\bf A}-\lambda {\rm\bf I})$. 
The structure of the graph $\mathcal{G}$ associated to ${\rm\bf A}-\lambda {\rm\bf I}$ reveals the symbols contributing to $\chi$. 
One can then determine which changes of off-diagonal entries leave all contributing symbols and thus the spectrum invariant.

%%% CONCLUSIONS %%%
In the present Letter, we proposed a graphical notation that facilitates the computation of spectral properties of complex systems. 
Applying this notation to systems of symmetrically-coupled one-dimensional dynamical units, we showed that any system obeying 
a simple force balance condition (ZRS) has to obey a global stability condition: Local dynamical stability of steady (e.g.{} phase-locked) states requires that the Coates graph of the Jacobian has a spanning tree of positive links. This criterion is complementary to results obtained by other methods (master stability function, ensemble simulations) and pertains to a large class of systems studied in physics, containing the Kuramoto model. Along with similar rules that can be derived analogously, the spanning-tree criterion has distinct meso-scale consequences, limiting for instance the number, position, and strength of negative elements in network motifs. Beyond the coupled oscillators, the proposed approach is applicable to questions ranging from adaptive networks in neuroscience to isospectrality problems in condensed matter.      

We are grateful to Jeremias Epperlein and Stefan Siegmund for valuable discussions.
%%%%%%%%%%%%%%%%%%%%%%%%%%%%%%%%%%%%%%%%%%%%%%%%%%%%

\end{document}